# A mechanistic-statistical species distribution model to explain and forecast wolf (*Canis lupus*) colonization in South-Eastern France

Julie Louvrier[1,2], Julien Papaïx[3], Christophe Duchamp[2], Olivier Gimenez[1]

[1] CEFE, CNRS, Univ Montpellier, Univ Paul Valéry Montpellier 3, EPHE, IRD, Montpellier, France

[2] Office Français de la Biodiversité, Unité Prédateurs Animaux Déprédateurs et Exotiques, Parc Micropolis, 05000 Gap, France

[3] BioSP, INRA, Avignon, France

Corresponding author: Julie Louvrier, CEFE UMR 5175, CNRS, Université de Montpellier, 12 Université Paul-Valéry Montpellier, EPHE, Montpellier Cedex 5, France, 13

julie.louvrier2@gmail.com

julien.papaix@inra.fr

christophe.duchamp@ofb.gouv.fr

olivier.gimenez@cefe.cnrs.fr

**Abstract**

Species distribution models (SDMs) are important statistical tools for ecologists to understand and predict species range. However, standard SDMs do not explicitly incorporate dynamic processes like dispersal. This limitation may lead to bias in inference about species distribution.

Here, we adopt the theory of ecological diffusion that has recently been introduced in statistical ecology to incorporate spatio-temporal processes in ecological models. As a case study, we considered the wolf (*Canis lupus*) that has been recolonizing Eastern France naturally through dispersal from the Apennines since the early 90's. Using partial differential equations for modelling species diffusion and growth in a fragmented landscape, we develop a mechanistic-statistical spatio-temporal model accounting for ecological diffusion, logistic growth and imperfect species detection. We conduct a simulation study and show the ability of our model to i) estimate ecological parameters in various situations with contrasted species detection probability and number of surveyed sites and ii) forecast the distribution into the future. We found that the growth rate of the wolf population in France was explained by the proportion of forest cover, that diffusion was influenced by human density and that species detectability increased with increasing survey effort. Using the parameters estimated from the 2007-2015 period, we then forecasted wolf distribution in 2016 and found good agreement with the actual detections made that year. Our approach may be useful for managing species that interact with human activities to anticipate potential conflicts.

1. **Introduction**

Assessing the dynamics of species distributions is a fundamental topic in ecology (Elith & Leathwick 2009). Species distribution models (SDMs) have become tremendously important tools in the fields of ecology, biogeography and conservation biology to understand and predict species distribution by correlating occurrence data to environmental covariates (Guisan & Thuiller 2005). SDMs can be used to study distribution dynamics through time (Elith & Leathwick 2009; Kéry et al. 2013; Hefley & Hooten 2016; Koshkina et al. 2017), which is especially relevant in conservation for the management of threatened species, conservation planning, as well as predicting the likely future range of invasive species at early invasion stages (Elith & Leathwick 2009; Guillera-arroita et al. 2015).

Despite being the most widely used methods in ecological applications, SDMs based on regressing presence locations on environmental factors suffer from several limitations (Hefley & Hooten 2016; Hefley et al. 2017b). These standard SDMs rely on the hypotheses that species will be present in the most favorable areas and that dispersal is not a limiting factor (Jeschke & Strayer 2006). However, expanding species may be absent from an area because they have not yet dispersed to this area, or because of geographical barriers or dispersal constraints (Araújo & Guisan 2006), not necessarily because conditions are unfavorable.

Species may expand through colonization defined as the ecological process of populations' establishment in unoccupied areas, in which populations can often face novel environments (Koontz et al. 2017). Colonization is therefore a dynamic process, underlying the past, present and future distribution of species (Clark et al. 2001; Wikle 2003; Wikle & Hooten 2010; Williams et al. 2017). Colonization can be a natural process, or the consequence of anthropogenic pressures, for example biological invasions (Sakai et al. 2001; Ricciardi 2007). Being able to understand the underlying mechanisms of the colonization has significant implications for wildlife managers (Koontz et al. 2017). Ignoring the dynamic process underlying distribution change can lead to biased inferences and some authors have discouraged the use of traditional, static SDMs for predictions (Yackulic et al. 2015).

Mechanistic spatio-temporal models have been developed to offer an alternative to regression-based SDMs that encounter difficulties associated colonization as a consequence of dispersal processes (Hefley et al. 2017b). Mechanistic models are based on biological processes, such as survival or dispersal, describing processes through which environmental factors affect a biological system of interest (Morin & Thuiller 2009; Mouquet et al. 2015; Gauthier et al. 2016). SDMs accounting for dynamic mechanisms are relevant tools to assess ecological colonization, because they improve our ability to get predictions in space and time and at the same time include reliable measures of prediction errors (Williams et al. 2017).

The theory of ecological diffusion is an essential component of mechanistic models to assess spatial distributions dynamics and population dynamics (Soubeyrand & Roques 2014; Roques & Bonnefon 2016; Hefley et al. 2017a, 2017b). To model dynamic ecological processes, mechanistic models are often expressed as partial differential equations (PDEs) (Wikle & Hooten 2010). Such PDEs can be combined with a probabilistic observation process in a mechanistic-statistical approach to infer biological sound parameters while considering complex observational protocols (presence only data, imperfect detection, censoring). In addition, combining a mechanistic-statistical model with a probabilistic observation process facilitates forecasting spatio-temporal processes (Wikle et al. 1998).

Here, we aimed at exploring the use of mechanistic-statistical models to gain insight into the colonization process of expanding populations of large carnivores, with a particular emphasis on an explicit modeling of the observation process that links the true states to the observed data. Indeed, data collection is particularly costly for elusive species that need wide areas to live and/or disperse. Monitoring large carnivores often requires sampling large areas. In this context, opportunistic data produced by semi-structured citizen science are increasingly used as an efficient source of information to assess the dynamics of such species (Schmeller et al. 2009; Louvrier et al. 2018; Kelling et al. 2019). The monitoring system often relies on the only available opportunistic data, leading to a set of presence locations, without any information about absences (Koshkina et al. 2017). These data need to be analyzed cautiously as they are collected without any measure of time- and space-varying sampling effort, possibly leading to biased estimates of the actual factors influencing the distribution (Van Strien et al. 2013). Furthermore, large carnivores can go undetected at sites where they are actually present, due to imperfect detection (Kéry 2011). Ignoring the issue of imperfect detectability of individuals can lead to underestimating the actual distribution (Kéry & Schaub 2011; Kéry et al. 2013; Lahoz-Monfort et al. 2014) and confounding between the environmental factors driving the distribution dynamics and those governing the observation process (Lahoz-Monfort et al. 2014).

Here, we developed a mechanistic-statistical model accounting for ecological diffusion, logistic growth and imperfect detection varying in space and time. The goals of our study were to i) provide a template to simulate scenarios and assess the ability of our method to reliably forecast the fate of populations in time and space and ii) provide an easy and convenient way to implement the approach in software heavily used by statisticians and ecologists such as JAGS and OpenBUGS.

To assess the performance of our approach, we performed a simulation study to assess bias and precision of parameter estimates and evaluate forecasting performance in contrasted scenarios of varying species-level detectability and number of monitoring sites. Finally, we fitted our model to opportunistic data on wolves in South-Eastern France between over nine years (2007-2015). We considered grey wolves (*Canis lupus*) as a case study to illustrate the challenges of using detections/non-detection data to infer the dynamics of a recolonizing large carnivore population. Wolves disappeared in western European countries during the twentieth century (Mech & Boitani 2010; Chapron et al. 2014) except for Spain, Portugal and Italy (Ciucci et al. 2009). The species then naturally recolonized the French Alps from the remnant Italian population (Valière et al. 2003). Starting in the 1990s, the species then spread outside the Alpine mountains to reach the Pyrenees and the Massif Central then later, even the Vosges Mountains in the North. In areas with livestock farming, conflicts may arise between wolf presence and sheep breeding. Because wolves are protected by law while being a source of conflicts with shepherds, their recolonization process needs to be carefully monitored. Besides quantifying the wolf colonization process over the study period, we explored the ability of our model for short-term forecasts of wolf range expansion.

## 2. Material and Methods

### 2.1. Model

We developed an approach to infer the parameters from a mathematical formulation explaining the temporal dynamics of the species' distribution (see also Hooten and Hefley 2019, chapter 28). To do so, we adopted the framework of ecological diffusion (Turchin 1998; Hefley et al. 2017b). We developed a state-space modelling approach in which the model is formulated in two parts: 1) the observation process that handles the stochasticity in the detections and non-detections (i.e., the observed distribution data) conditional on 2) the latent state process which is described by the mechanistic model.

#### 2.1.1. Observation process

Let $y_{ijt}$ be a random variable that takes value 1 if at least one individual is detected at site $i = 1,..., K$ at site $i$ within a study area S ($i \in S \subset R^2$) during secondary occasion (or survey, defined as a repeated sampling occasion during which the states of a site $i$ remains constant) $j = 1,..., J$ in year $t = 1,..., T$, and takes value 0 otherwise. Let $N_{i,t}$ be the true abundance at site $i$ in year $t$. The probability $p_{it}$ for the species to be detected at site $i$ in year $t$ is likely to be influenced by abundance $N_{it}$. To link the detection process to abundance, we used the Royle-Nichols formulation (Royle & Nichols 2003) developed to deal with heterogeneity in the detection probability due to variation in abundance and/or surveys (Williams et al. 2017). If at a site $i$ during year $t$ there are $N_{it}$ individuals present, assuming that each individual within an occupied site has an identical detection probability $q_{it}$, and that there is independence of detections among individuals, then the probability to detect the species is equal to the probability to detect at least one of the $N_{it}$ individuals present. This latter probability is the complementary probability of failing to detect any individual and can be written as $(1 - q_{it})^{N_{it}}$. Therefore, the probability to detect at least one individual at site $i$ during year $t$ can be written as follows:

$$p_{it} = 1 - (1 - q_{it})^{N_{it}}. \qquad (1)$$

Conditioning the observation $y_{i,j,t}$ on the latent, true abundance $N_{it}$ through the species-level detection probability $p_{it}$, and assuming a binomial observation process, a constant survey effort, and that $q_{it}$ and $N_{it}$ remain unchanged across the $J$ surveys, we then have

$$y_{it} = \sum_{j=1}^{J} y_{ijt} \sim \text{Binomial}(J, p_{it}). \qquad (2)$$

The $J$ repeated surveys within each year $t$ are used to estimate the species-level detection probability. Note that if $N_{it} = 0$ then $p_{it} = 0$ and $y_{ijt} = 0$ for all $j$.

Covariates may be incorporated in the individual-level detection probability $q_{i,t}$ using, for example, a logit link function. Because we had information about the sampling effort, sites that were considered sampled were sites where sampling effort was non-null. To the contrary, sites that were considered as non-sampled (i.e. on which no information about detection can be made) were sites with a sampling effort equal to zero. To avoid estimating the detection probability where sampling effort was null, we set the detection probability to zero when sampling effort was equal to zero.

### 2.1.2. State process

We assumed that the true abundance $N_{i,t}$ at site $i$ during year $t$ was Poisson distributed over a site $i$

$$\begin{cases} N_{it} \sim \text{Poisson}(\lambda(i,t) \times \epsilon_{it}) \\ \log(\epsilon_{it}) \sim \text{Normal}(0,\sigma) \end{cases}, \quad (3)$$

where $\epsilon_{i,t}$ accounts for overdispersion. The variable $\lambda(i,t)$ is a spatiotemporal process that describes the dynamics of the number of individuals in site *i* during year *t*. We then defined this variable as follows:

$$\lambda(i,t) = \int_{B_i} v(x,t)dx, \quad (4)$$

where *v(x,t)* is the intensity of individuals at the spatial location *x* at time *t* and $B_i$ is the study area in which counts occur.

We used Partial Differential Equations (PDE) known as ecological diffusion to describe diffusion and growth dynamics. The ecological diffusion PDE describing the variation of density of individuals at location *x* at time *t*, *v(x,t)* over time, in two dimensions with logistic growth (see also Lu et al. 2019), can be written as follows:

$$\frac{\partial v(x,t)}{\partial t} = \Delta(d(x)\,v(x,t)) + r(x)\,v(x,t)\left(1 - \frac{v(x,t)}{K}\right), \quad (5)$$

where $\Delta$ is the Laplace 2D diffusion operator (i.e. the sum of the second derivatives with respect to the coordinates). This operator describes movement according to an uncorrelated random walk, with the coefficient *d(x)* measuring heterogeneous mobility. The term *r(x)* is the intrinsic growth rate at low density and *K* is the carrying capacity. In short, this equations states that the variation of density of individuals at a location *x* at time *t* is the result of a diffusion process and a logistic growth process. The diffusion process is governed by an inflow of individuals diffusing from the neighboring cells and an outflow of individuals diffusing to the neighboring cells, with *d(x)* accounting for the heterogeneity of these diffusion flows (Hefley et al. 2017b; Williams et al. 2019). The logistic growth process is governed by a logistic growth parameter *r(x)*, defined as the rate of increase of a population at site *x*, and *K* the carrying capacity, defined as the maximum number of individuals a site can sustain indefinitely. To fit our model, we made some assumptions about the parametric distributions about these parameters, which can be found in sections "Simulations" and "Case study". In addition, we assumed reflecting boundary conditions, meaning that there was no population flow going outside the boundaries of the study area due to actual barriers (i.e. seas) or symmetric inward and outward flows.

*2.1.3 Approximations*

Calculating the density *v*(x,t) requires solving the PDE described in equation 5. We used the method of lines (Schiesser 1991; Chow 2003) to approximate the PDE by a system of Ordinary Differential Equations (ODE) in order to use classical numerical integration algorithm to solve the dynamical system. The methods of lines consist in discretizing the spatial domain into $C_s$ grid cells of *O* rows and *L* columns leading to the following ODE system, with $u(i,t)$ the discretized approximation of *v*(*x,t*) at site *i*:

$$\dot{U}_t = R \times U_t \left(1 - \frac{U_t}{K}\right) + M\, U_t, \qquad (6)$$

where $U_t^T = [u(1,t), u(2,t), ..., u(C_s,t)]$ is the vector of densities in each cell, $R^T = [\bar{r}(1), \bar{r}(2), ..., \bar{r}(C_s)]$ is the vector of averaged intrinsic growth rates in each cell and × indicates the term by term product. M is the $C_s$ x $C_s$ propagator matrix that describes spatial interactions between direct neighboring cells in the four cardinal directions. The $i$th row of M represents the link between the $C_s$ sites to site *i*. The approximation of the differential operator in equation 5 is then:

$$[MU_t]_{s_{k,l}} = \frac{1}{h^2} \left[ d(s_{k+1,l})u(s_{k+1,l}, t) + d(s_{k-1,l})u(s_{k-1,l}, t) + d(s_{k,l+1})u(s_{k,l+1}, t) + \right.$$
$$\left. d(s_{k,l-1})u(s_{k,l-1}, t) - 4d(s_{k,l})u(s_{k,l}, t) \right], \quad (7)$$

with $s_{k,l}$ the coordinates of the site $i$, i.e. $s_{k,l} = l(k-1) + l$ ; $h^2$ the cell surface ; $k = 1,..., O$; $l = 1,..., L$ and $O \times L = C_s$. Exceptions are the cells bordering non-habitat cells as the latter are excluded from the dynamics due to the reflecting boundary conditions. The system 6 was solved using the lsoda method (Petzold 1983) through the R-package deSolve (Soetaert et al. 2010) and equation 4 was then approximated as follow:

$$\lambda(i, t) = \int_{B_i} v(x, t) dx \approx \sum_{k=1}^{O} \sum_{l=1}^{L} \mathcal{A}(B_i \cap c_{s(k,l)}) u(s_{k,l}, t), \quad (8)$$

where $\mathcal{A}(B_i \cap c_{s(k,l)})$ is the surface of the intersection between the cell $s(k, l)$ and the study area $B_i$ in which counts occur.

## 2.2. Simulations

We conducted a simulation study to assess the ability of the model to estimate ecological parameters. We defined four scenarios in which we explored the effect of a variation in the grid resolution (see section *Approximations* above) and in the individual-level detectability parameter $q$. To mimic the characteristics of the wolf case study (see below), we set the number of surveys to 4 and the number of years to 20, while we set the carrying capacity to 5 individuals per 100 km², the intercept of the diffusion coefficient to 2 individuals per cell (i.e. 5 individuals per year per cell move to neighboring cells) and the growth rate to 40%. We set the linear and quadratic effects of forest cover on the growth rate at 0.4 and 0.4 and set the linear and quadratic effect of human density on the diffusion rate at 0.5 and 0.3 respectively. We randomly simulated values of forest density and human density between their maximum and minimum values from the wolf study. Because we discretize the spatial domain, we expected a lower bias and a better precision of the ecological parameters estimates with increasing grid cell resolution. We defined a low resolution (LR) scenario in which the spatial domain to fit the model was divided into 25 cells and a high resolution (HR) scenario in which we divided the spatial domain into 100 cells and fitted the model to this resolution. In both scenarios, we simulated the ecological data on a grid of 100 sites resolution. Under the Royle-Nichols formulation of the relationship between abundance and binary detection and non-detection data, individual-level detectability has a positive effect on the species-level detectability until a certain level of abundance, hence it influences whether the species is detected or not. We then defined a high detectability (HD) scenario in which the individual-level detectability was set at 0.8, and a low detectability (LD) scenario in which this probability was set at 0.2. For each scenario (LR-LD, LR-HD, HR-LD, HR-HD), we simulated 100 datasets and we fitted the model to each dataset. We calculated the relative bias and mean squared error (MSE) for the carrying capacity $K$, the intercept of the growth rate $R$, the linear and quadratic effect of forest density on the growth rate, the diffusion coefficient and the linear and quadratic effect of human density on the diffusion coefficient. Note that in the simulation study we assumed that $K$, $R$ and $q$ were constant over space and time. To explore the ability of our model to forecast the abundance of individuals per site in the four scenarios, we fitted our model to the first ten years and forecasted the distribution over second ten years.

### 2.3. Case study: Wolf colonization in France 2007-2015

Wolf detections and non-detections were made in the form of presence signs sampled all year round in a network of widely distributed professional and non-professional wolf observers (Duchamp et al. 2012). Presence signs went through a standardized control process to prevent misidentification.

To define the observation process, we used a grid to cover the study area with 10x10 km cells that we used as sampling units ($C_s$ = 975 in the notation above). To ensure that the model we fitted to this discretization choice produces reliable estimates, we estimated the parameters based on a 3x3km grid. We then simulated the dynamic model with the estimated parameters and calculated the relative error (RMSE) in comparison with the finest grid. We found that a resolution of 10x10 km produced a relatively low error in comparison with a finer grid size (Appendix 1).

Wolf monitoring occurred mainly in winter from December to March, the most favorable period to detect the species. Within each winter, four secondary occasions were defined as December, January, February and March (i.e., $J = 4$). We focused on the south-eastern part of France and the period 2007-2015 ($T = 9$) (Fig. 1). We assumed that the scale at which data were collected coincides with the numerical scale in which we solve $u(i,t)$, thus equation 8 becomes $\lambda(i,t) \approx h^2 u(i,t)$. We used the sampling effort, defined as the number of observers at site $i$ in year $t$ (Eff$_{it}$) and the road density at site $i$ (RoadD$_i$) to explain variation in the individual-level detection probability ($q_{i,t}$) as:

$$\text{logit}(q_{it}) = \beta_0 + \beta_1 \text{Eff}_{it} + \beta_2 \text{RoadD}_i. \tag{9}$$

We expected that the sampling effort had a positive effect and road density had a negative effect on the individual-level detection probability $q$. We also used environmental and anthropogenic covariates to model spatial variation in parameters $R_i$ and $D_i$. Using the CORINE Land Cover® database (U.E – SOeS, Corine Land Cover 2006), we calculated forest cover as the average percentage of mixed, coniferous or deciduous forest cover for each site. Because forests may structure the ungulate distribution (i.e. prey species), we expected that forest cover would have a positive effect on the logistic growth rate $R_i$ (Louvrier et al. 2018).

Human density was found in previous studies to influence habitat choice and dispersal of wolves in Italy (Corsi et al. 1999; Marucco & Mcintire 2010). We therefore considered human density as a candidate covariate possibly explaining spatial variation in the diffusion parameter $D_i$. Human population density was averaged in each 10x10 km from a 1x1 km raster from the Earth Observing System Data and Information System (EOSDIS). For both parameters, we tested a linear and a quadratic effect through a logarithmic, for $D_i$, and a logistic limited between 0 and 2, for $R_i$, regression-type relationship.

Finally, we initialized the model with $\lambda = 0.01$ for the sites with at least one wolf detection during the period 1994-2006 preceding our study period, which corresponds to one individual per 100 km² cell, and zero otherwise.

To explore the ability of our model to forecast wolf colonization over the short term, we used the parameter estimates we obtained on the 2007-2015 period and forecasted the abundance one year ahead (i.e., to 2016). We assessed our predictions qualitatively by confronting the estimated probability of a site being occupied (forecasted abundance at that site > 0) in 2016 to the observed detections made in that same year.

### 2.4. Bayesian inference

To complete the Bayesian specification of our model, we specified Gaussian priors with mean 0 and variance 1 for all estimated parameters, except for parameter $K$ for which we used a logistic function limited between 0 and 0.2. Parameters from the observation process and those from the state process were updated in two different blocs. We implemented our simulations in OpenBUGS (Lunn et al. 2010) and the wolf analyses in JAGS using the JAGS package mecastat (Rey et al. 2018). We used Markov chain Monte Carlo (MCMC) simulations for parameter estimation and forecasting (Hobbs & Hooten 2015). We ran three MCMC chains with 40,000 iterations preceded by 10,000 iterations as a burn-in. We used posterior medians and 95% credible intervals to summarize parameter posterior distributions. We checked convergence visually by inspecting the chains and by checking that the R-hat statistic was below 1.1 (Gelman & Shirley 2011). We produced distribution maps of the latent states by using a posteriori means of the $N_{i,t}$ from the model. We provide the scripts for running the simulations at https://github.com/oliviergimenez/appendix_mecastat.

### 2.5. Forecasting

To forecast the abundance of individuals per site, along with the associated prediction uncertainty, we needed to assess the probability distribution of the true state in the future when data will be collected, conditional on the collected data in the past (Williams et al. 2018). The Bayesian formulation of our model allowed assessing the forecast densities by simulating yearly data from $t = 2, …, T + 1$ and sampling $\lambda(i, T+1)$ on each iteration of the MCMC chains. The posterior distribution was then assessed from the forecast distribution by sampling into the forecast $N_{T+1}$. In the simulation study, we compared this posterior distribution with the simulated data for the year 20. For the wolf case study, we assessed the probability that the site $i$ was occupied, which boiled down to calculating $P(z_i = 1)$ where $z_i$ is the latent status of the site (occupied or not) as the number of MCMC iterations producing a strictly positive abundance, i.e. $P(z_i = 1) = P(N_i > 0)$ (since our distribution model is formulated in terms of latent abundance $N$).

## 3. Results
### 3.1. Simulations

When the resolution from which we fitted the model increased from 25 cells to 100 cells, the model produced less biased results for all parameters, except the linear and quadratic effects of human density on the diffusion coefficient (Fig. 2 and Appendix 2. A.). For the carrying capacity the bias decreased from -6.09 % in LR-HD and -1.91 % in LR-LD and only 1.57 % in HR-HD and 0.70 % in HR-LD. The bias also decreased for the intercept of the growth rate when resolution increased: - 66.63 % in LR-HD and -64.89 % in LR-LD to 10.54 % in HR-HD and 11.94 % in HR-LD. For the intercept of the diffusion coefficient, the bias was reduced from -25.62 % in LR-HD, -9.95 % in LR-LD and 1.43 % in HR-LD to 3.67 % in HR-HD.

The model also produced more precise results for all parameters, except the linear and quadratic effects of human density on the diffusion coefficient (Fig. 2 and Appendix 2. A.). The largest MSE reduction was found for the carrying capacity. The MSE decreased for the carrying capacity from 1.89 in LR-HD and 0.80 in LR-LD to 0.22 in HR-HD and 0.21 in HR-LD. For the intercept of the diffusion coefficient the MSE decreased greatly from 0.43 in LR-HD and 0.34 in LR-LD to 0.06 in HR-HD and 0.01 in HR-LD. We didn't find any clear pattern in the change of MSE for the growth rate. In both high and low detectability scenarios, the model fitted in low resolution largely underestimated the linear and quadratic effects of forest density on the growth rate.

Without covariates on the diffusion parameter and the growth rate, when the resolution increased the model produced less biased and more precise results as well (Appendix 2.B. and 2.C.)

When looking at the model's ability to forecast abundance (Appendix 3), the true abundance was always within the 95 % credible interval of the estimated abundance in both high resolution scenarios and in the low resolution high detectability, but not in the low resolution low detectability scenario.

### 3.2. Wolf case study

According to our model, the estimated abundance per site varied between 0 and 19 per 100 km$^2$ (Fig. 3, Appendix 4 for the credible intervals. Overall, the spatio-temporal trends in estimated abundance matched relatively well the trends in actual detections and non-detections (Fig. 3).

In the northern part of the study area, we estimated a non-null abundance at sites where no detections were made in the last four years of the study.

The detection probability increased when the sampling effort increased and decreased when road density increased (Fig. 4 and Appendix 5). We found that the logistic growth rate increased when the forest cover increased. The carrying capacity was estimated around 1 individual per 100 km$^2$ site (9.41x10$^{-3}$ CRI: 7.97x10$^{-3}$; 1.11x10$^{-2}$). Last, when human density increased, the diffusion parameter increased as well.

Turning to the forecasting exercise now, we predicted a median abundance varying between 0 and 1 individual per site, while the 95% credibility interval predicted an abundance varying between 0 and 17 individuals per site (Appendix 6). For the forecasted occupancy, we predicted that a large part of sites with a forecasted occupancy probability > 0.6 were indeed detected occupied in year 2016 (Fig. 5). Amongst the 137 sites that were detected occupied in 2016, we found only 10 of them in the South-Western part which were forecasted with a low occupancy probability. This leads to a false negative rate of 7.30%. However, the model forecasted a higher number of sites with a high occupancy probability than the number of detected occupied sites.

## 4. Discussion

We estimated the distribution of wolves using a model explicitly incorporating biological mechanisms and making best use of the information contained in species detections and non-detections. Besides, we explored the possibility of forecasting the potential future distribution of a large carnivore, which could be used to target management areas or focus on potential conflictual areas (Marucco & Mcintire 2010; Eriksson & Dalerum 2018).

### 4.1. Simulations

In the simulation study, we showed that ecological parameters were sensitive to the way we discretized space to solve the PDE. Our model performed well when the resolution was high, with less biased and more precise parameter estimates than in the low-resolution scenario. We note however that the low-resolution scenario was an unrealistic example used to test the model in extreme conditions.

### 4.2. Wolf study

We found that the logistic growth rate increased when forest cover increased. Although wolves can adapt to various ecosystems, this pattern also matches with the suitable habitats of the key prey species for wolves (Darmon et al. 2012). We found that when human density increased, the diffusion coefficient increased possibly due to the increase of linear features, which have been found to be selected over natural linear features for wolves' movements (Newton et al. 2017).

As expected, we found that when sampling effort increased, the individual-level detectability increased, while it decreased when road density increased. We also expected that road density would influence wolf detectability by facilitating observers survey and site accessibility. Other studies have found that linear features also facilitate wolf travel and prey encounter rate. On the contrary, we found that the increase in road density negatively affected the species detection. This result was found in previous studies as well corroborating the fact that wolves avoid roads and leave fewer marks in sites with highly frequented roads or pathways (Whittington et al. 2005; Falcucci et al. 2013; Votsi et al. 2016; Louvrier et al. 2018).

We estimated a maximum number of 19 individuals per grid cell on average. Wolves pack size was documented on average at 3.8 individuals per pack in France (Duchamp et al, 2012) varying from 2 to a dozen individuals. Considering the average wolf territory size commonly reported between 100 and 400 km² in western and central Europe (Ciucci et al. 2009; Mech & Boitani 2010; Duchamp et al. 2012), our estimate overestimated the standard range of wolf densities reported elsewhere (Mech & Boitani for a review). The fact that we found a non-null abundance at sites in the northern part of the study area could be explained by the fact that in the Western and Southern part of the study area, the human density is at its highest values, with two of the three most important cities in France, Lyon and Marseille that are found in the Western and Southern part of the study area respectively. The model accounted for the imperfect detection and estimated those sites with a non-null abundance despite the fact that no detection was made. This also explains the number of forecasted occupied sites higher than observed.

### 4.3. Model Assumptions

We built our model based on several assumptions that need to be discussed. We assumed that the sampling effort was constant across surveys and that the individual-level detectability and the local abundance remained unchanged. First, it is likely that the sampling effort varies between surveys (months) due to human factors. However, we could only quantify the sampling effort between years, and had no information at the month level. Second, it is also likely that the individual-level detectability varies between months partly due to the varying sampling effort between months, but also to environmental conditions, such a snow cover represented by the month of survey (Louvrier et al. 2018). Third, the local abundance at a site is also likely to change between surveys. The choice to consider the wintering data survey, during which the social organization of packs is the most stable (Mech & Boitani 2010), may prevent a large part of this sampling heterogeneity according the sampling protocol implemented in the Alps by the wolf network (Duchamp et al, 2012). However, we cannot exclude that mortality or movements occur inside or outside the sites. In this case, the estimates for local abundance can be overestimated as the same individuals can be detected in two neighboring sites for instance., The distribution should in any case be interpreted cautiously and considered as area of use (MacKenzie 2006).

Under the Royle-Nichols model, the species-level detectability is a function of the individual-level detectability, but the relationship between these two parameters is not always linear and depends on the abundance of individuals at a site. If abundance is large (i.e., above 50 individuals), then individuals can be detected during all surveys, and no variability in the species-level detectability will be observed, which leads to the inability to characterize the abundance distribution (Royle & Nichols 2003). Overall, the Royle-Nichols model was originally developed to deal with heterogeneity in the detection probability due to heterogeneity in abundance and its outputs should be interpreted cautiously. Finally, our approach was based on a logistic growth, but other forms of growth could be investigated. For example, a growth accounting for an Allee effect would be of particular relevance for wolves for which the probability of finding a mate decreases in areas with low density (Hurford et al. 2006).

Another assumption relies on the model construction considering the diffusion equally for all individuals. Wolves have a strong social organization in packs and future works may consider the social aggregation of individuals when modeling the dynamic of the wolf distribution (see for instance Lewis et al. 1997 and Potts & Lewis 2014)).

We need to highlight here the fact that our model was realistic only because we fitted it on data from the core distribution of wolves in France. However, if we had extended our model to the whole country, we would expect less realistic estimates due to the fact that wolves not only disperse at short distance but also at long distance, especially on colonization fronts (Mech and Boitani 2010). In Louvrier et al. (2018), we found that the number of observed occupied sites at long distance also influenced the probability for a site to be occupied. Our model was deterministic and if we were to extend our model to the whole country, we would need to account for stochasticity in events when the population is at low density (Hurford et al. 2006). To do so, we could assimilate the detections for a year in which long distance dispersal occurred and was not predicted by the model and use these data to initialize the model for the next year. Finally, when we calculated the values of the covariates, we used the mean for each grid of 10x10km. By doing so, we lost information at a finer scale. Based on the error measure we found when we approximated the model on a 10x10km scale we considered the loss of information to be within a tolerable range.

4.4. Comparison with dynamic site-occupancy models

In Louvrier et al. (2018), a dynamic site-occupancy model was fitted to the same dataset, at a national level and between 1994 and 2016. We found in this previous study that when forest cover was high, the probability for an unoccupied site to be colonized the year after increased as well. This can be related to the logistic growth rate parameter, because once a site is colonized, the population will start growing. We found the same effects of sampling effort and road density on the species-level detectability, which can be explained by the fact that maximum abundance per site is low enough to guarantee a linear correspondence between species- detectability and individual-level detectability. In comparison with the map of occupancy estimated with a dynamic site occupancy model (top right panel of Figure 7 in Louvrier et al. 2018), we found that the mechanistic approach predicted more sites with an average occupancy probability > 0.6 than the dynamic site-occupancy model. The latter approach estimated a smaller number of occupied sites. This difference could be explained by the fact that occupancy models are regression-type models, which means that the estimated occupancy is linked to the data, while our mechanistic approach is based on a continuous model over time, which allows the spreading of individuals over several sites without having to be detected. Another explanation could be that we assumed a Poisson distribution for the number of individuals per site in our mechanistic model. A first way to overcome this problem is to use a negative binomial distribution to relax the constraint of equal mean and variance inherent to the Poisson distribution (White & Bennetts 1996). Another approach would be to directly model the dependence between individuals by explaining the pack structure in the mechanistic part of our model (Lewis et al. 1997).

4.5. Forecasting capacities

In the current context of fast-changing environments, predicting the future distribution or abundance of species is an increasing challenge in the field of ecology, where ecologists are calling for a more "predictive ecology" (Mouquet et al. 2015; Dietze 2017; Houlahan et al. 2017; Dietze et al. 2018; Maris et al. 2018). Ecological forecasting is the process of predicting the state of an ecological system with fully specified uncertainties (Clark et al. 2001). Forecasts should therefore be probabilistic (Gneiting & Katzfuss 2014; Dietze & Lynch 2019) to provide reliable uncertainties. Not accounting for uncertainties associated with predictions of future change in distributions can lead to misguided decisions by policymakers or managers (Gauthier et al. 2016). The Bayesian method provides a natural framework for making probabilistic forecasts because it easily handles uncertainty and variability in all components of a statistical model (Hefley et al. 2017b). We demonstrated using simulations that our model had satisfying forecasting capabilities. When we applied our approach to the wolf, we produced satisfying forecasts for the presence of wolves. In 2016, 137 sites were detected as being occupied, out of which 10 sites were not forecasted as occupied by our model. These sites were found at the edge of the distribution core in the South-Western part of the study site. This part of the distribution was recently colonized by wolves with first detections of wolves occurred there in 2014 and 2015 for the first time. Wolves are highly flexible and can live in various areas from maize cultures to high mountains (Kaczensky et al. 2012). This South-Western part are places where forest cover is lower and human density is higher than in the Alpine range. In the future we might expect the effects of forest cover to be weaker as wolves expand in different landscapes.

## 5. Conclusion

Mechanistic-statistical models are valuable tools to bring insight into the dynamic of species distribution. However, ecologists are often faced with cryptic species with detectability less than one. Here we developed a mechanistic-statistical model accounting for imperfect detection for wolf management in France. The model is flexible and can be used in a variety of contexts to assess the dynamic of species distribution by amending the observation process if required. By forecasting the distribution of wolves in France, we illustrate that our approach may provide a new tool in the context of the management of a species with possible conflictual interactions with human activities. Our approach resonates with the adaptive management framework where ecologists need to make decisions based on yearly estimates of population abundance and distribution (Marescot et al. 2013).

**Acknowledgements**

We gratefully acknowledge the help of people who have collaborated with the wolf monitoring network supervised by the French game and wildlife agency (ONCFS). We thank Lionel Roques and Olivier Bonnefon for helping us building the model. We also warmly thank Marc Kéry for his constructive comments on the manuscript. JL is thankful to the GDR 3645 Ecologie Statistique. We thank the Montpellier University and the ONCFS for a PhD grant to the first author. This work was partly funded by a grant from CNRS and "Mission pour l'interdisciplinarité" through its "Osez l'interdisciplinarité call."

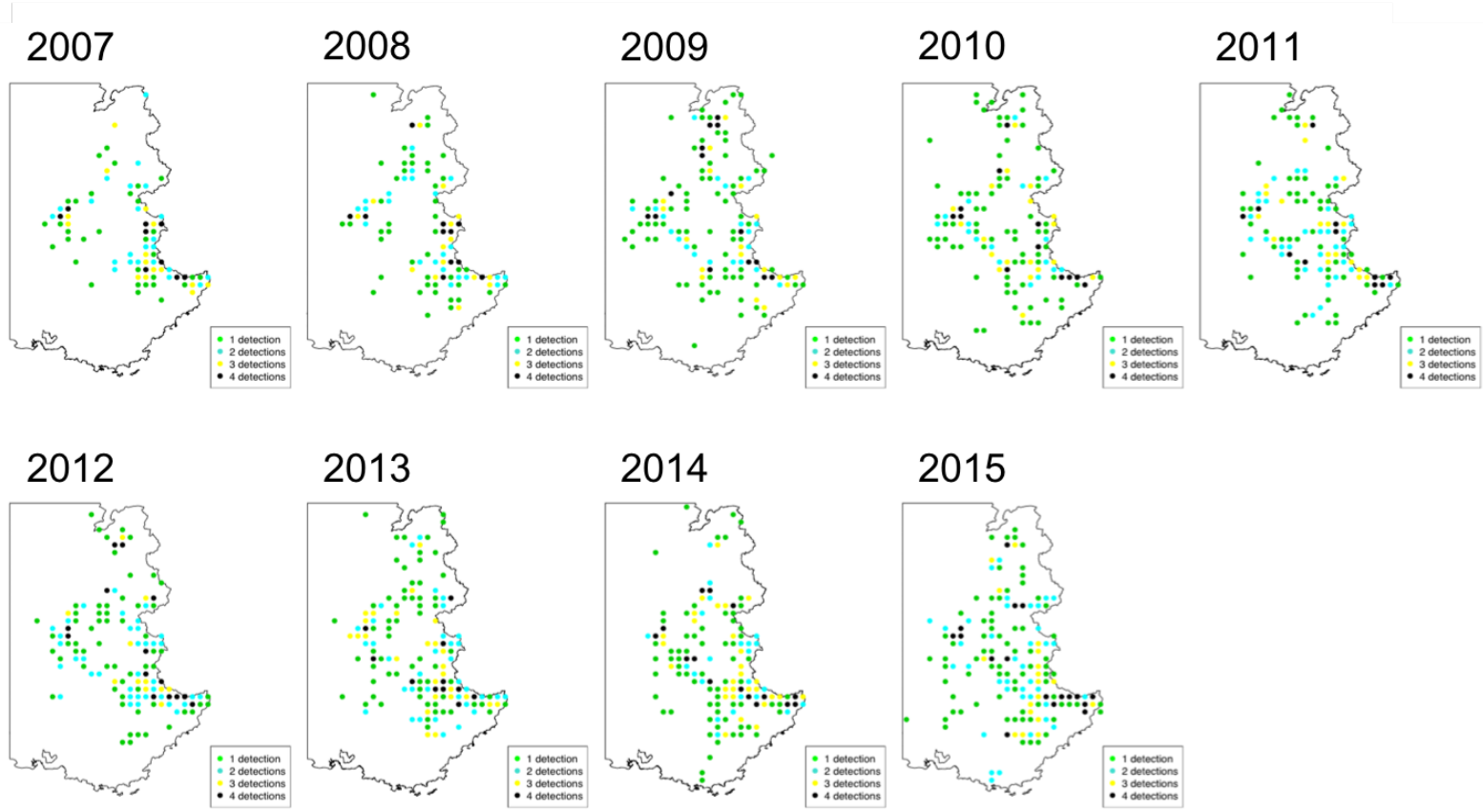

Figure 1: Maps of the yearly detections of wolf in the study area in France from years 2007 to 2015.

Figure 2: Performance of the model in the high resolution / high detectability scenario (left panels) and in the low resolution / high detectability scenario (right panels). For each of the 100 simulated datasets (on the Y-axis), we displayed the median (circle) and the 95% credible interval (horizontal solid line) of the parameter. The actual value of the parameter is given by the vertical dashed red line. The estimated bias (noted as "B") and MSE are provided in the legend of the X-axis.

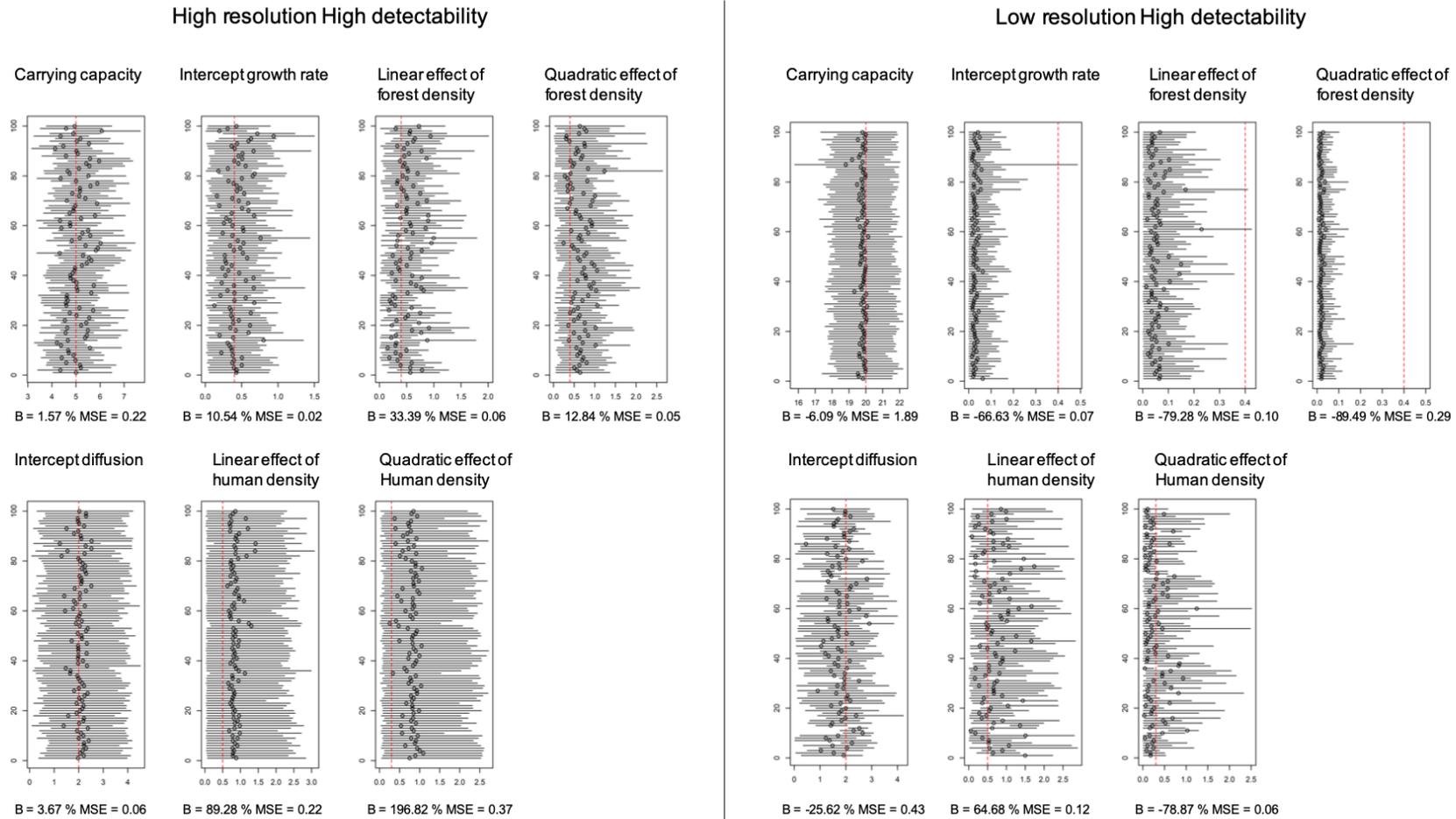

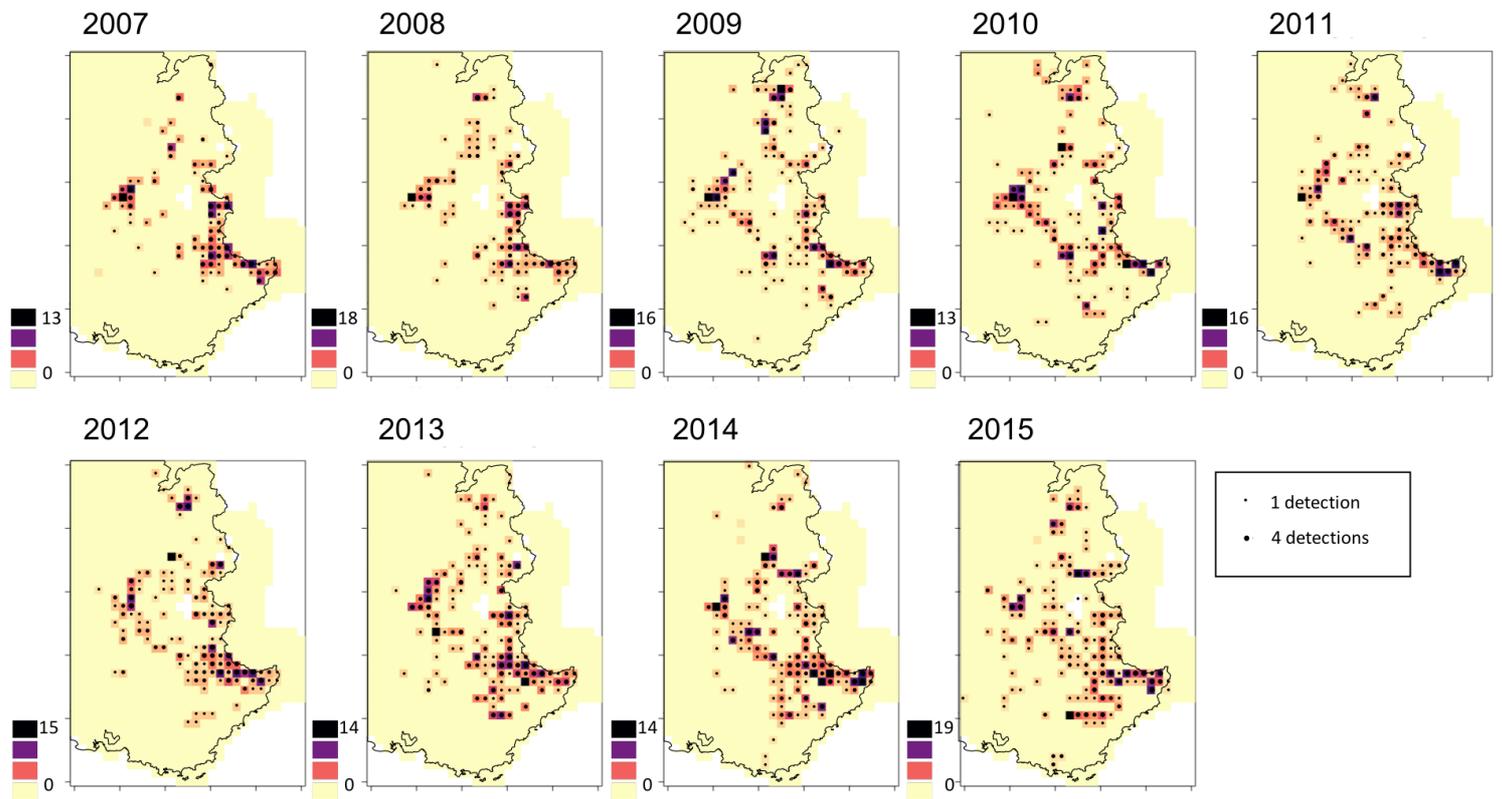

Figure 3: Maps of the estimated abundance of wolves per 100 km$^2$ site in South-East France between 2007 and 2015. Black dots represent detections in a year.

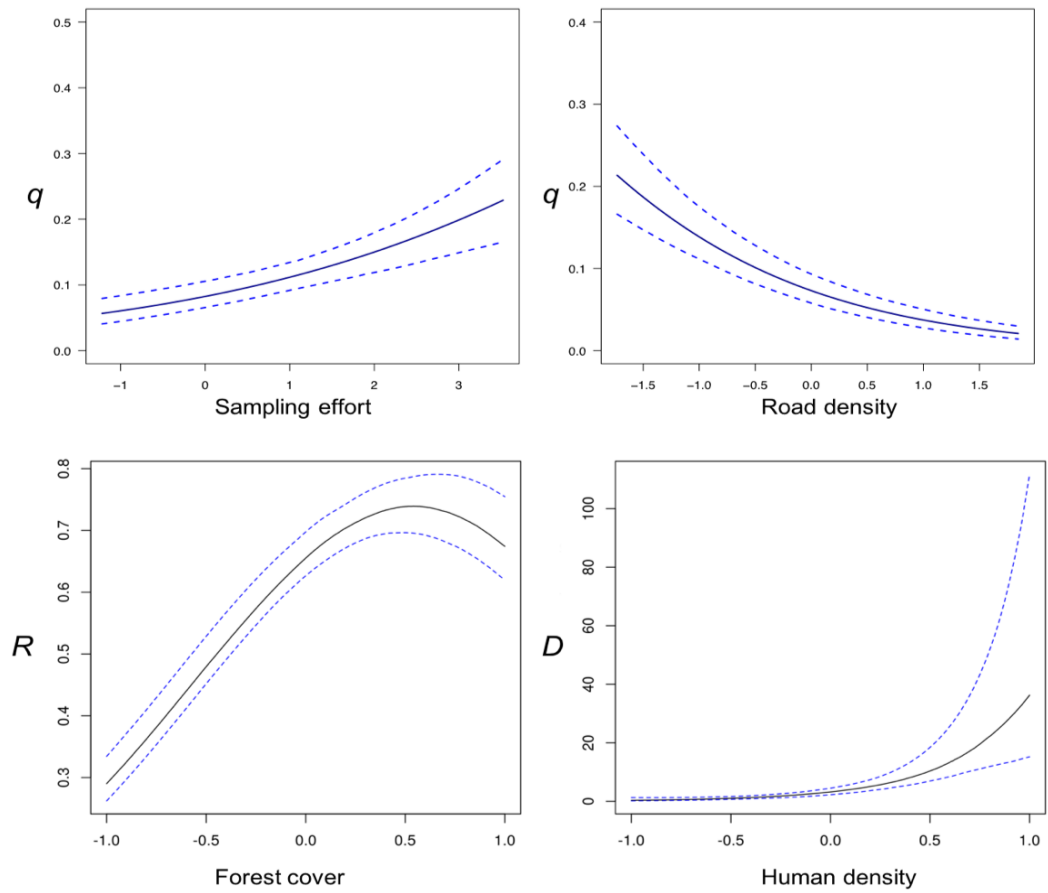

Figure 4: Estimated relationship between individual-level detectability and i) standardized sampling effort (top left) or ii) standardized road density (top right), between logistic growth rate and standardized forest cover (bottom left) and between diffusion and standardized human density (bottom right).

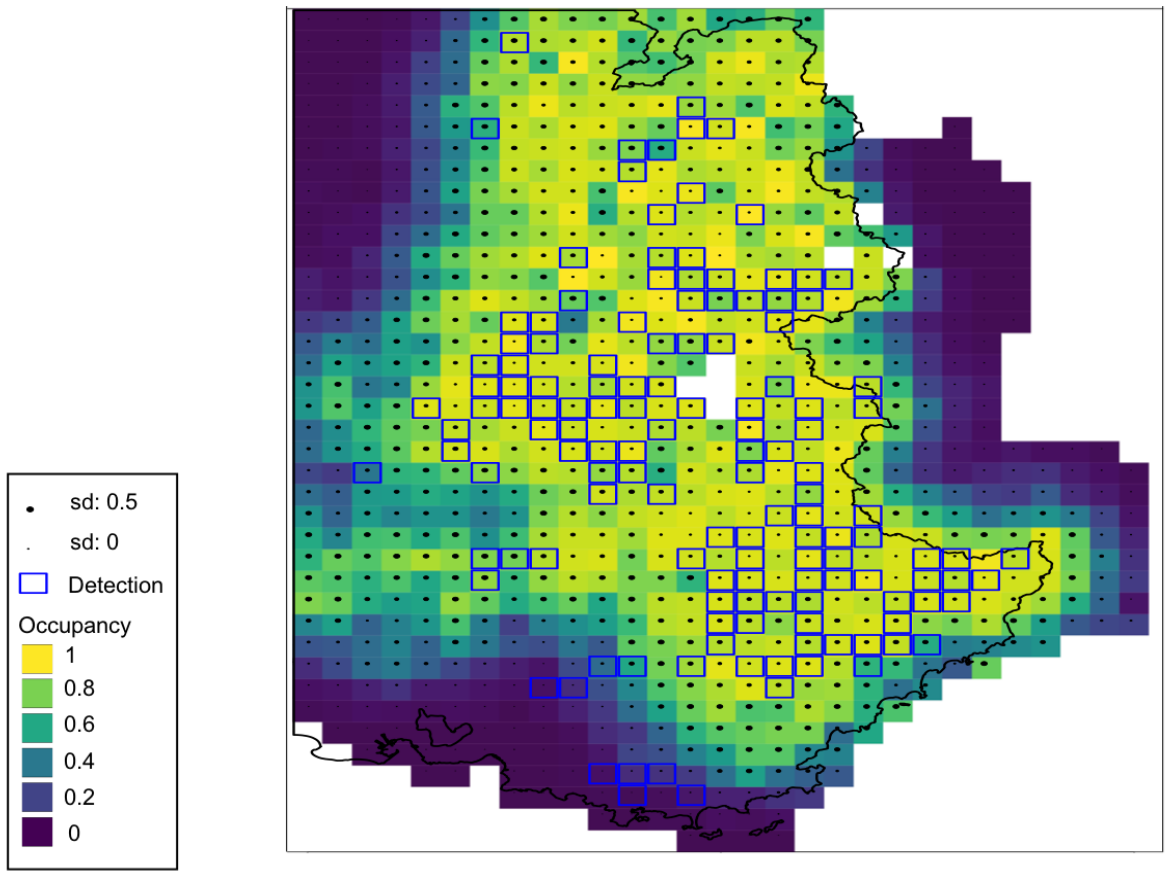

Figure 5: Map of the forecasted probability of occupancy for the year 2016 obtained from our mechanistic-statistical model fitted to the 2007-2015 period. The blue squares represent sites where detections occurred in 2016 and the black dots capture the prediction uncertainty, with the size of a black dot proportional to the standard deviation of the forecasted occupancy in the corresponding site (varying between 0 and 0.25).

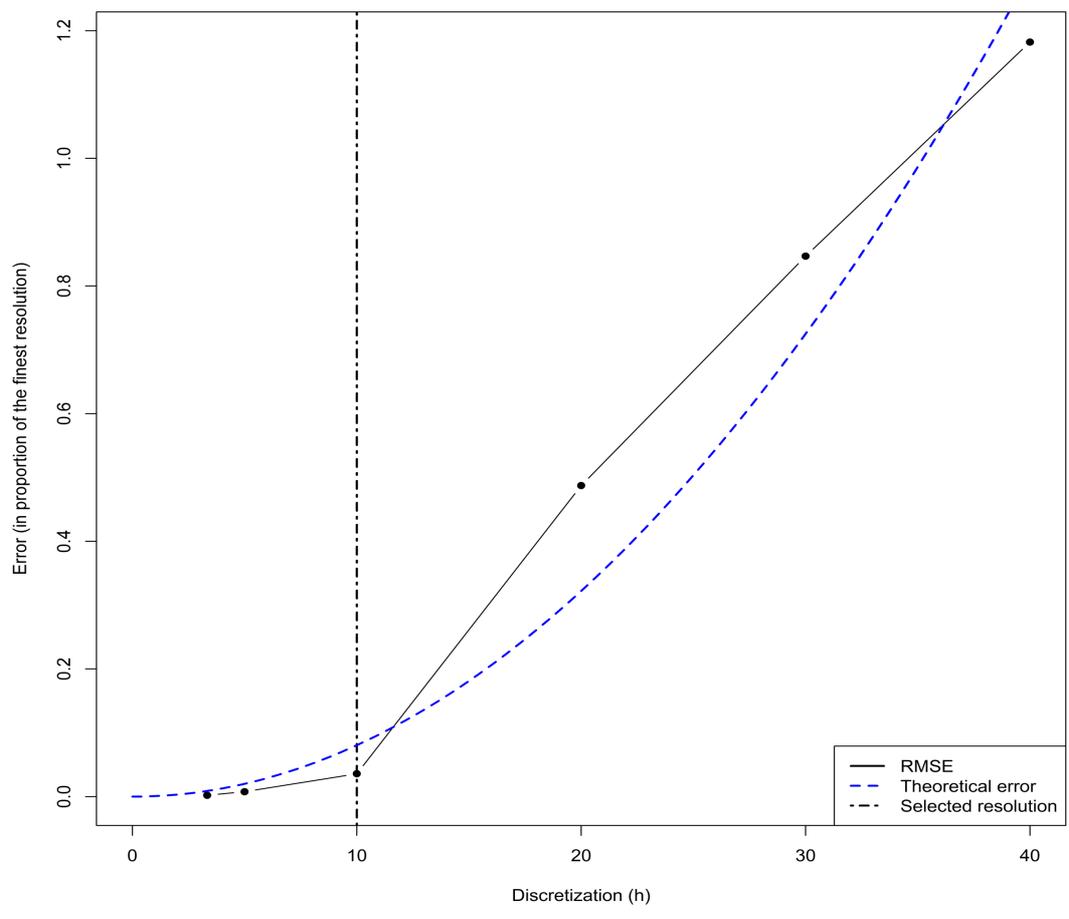

Appendix 1: RMSE of models fitted at different resolution, the RMSE was calculated in comparison with the estimates from the finest grid cell resolution defined as 3kmx3km. The Black line represents the observed error while the blue dotted line represents the theoretical error calculated as the quadratic term of the resolution. The black dotted line represents the resolution we chose for fitting our model on the wolf dataset.

10  Appendix 2: A. Performance of the model in the high resolution / low detectability scenario (left panels) and in the low resolution / low detectability
11  scenario (right panels). For each of the 100 simulated datasets (on the Y-axis), we displayed the median (circle) and the 95% credible interval
12  (horizontal solid line) of the parameter. The actual value of the parameter is given by the vertical dashed red line. The estimated bias and MSE are
13  provided in the legend of the X-axis

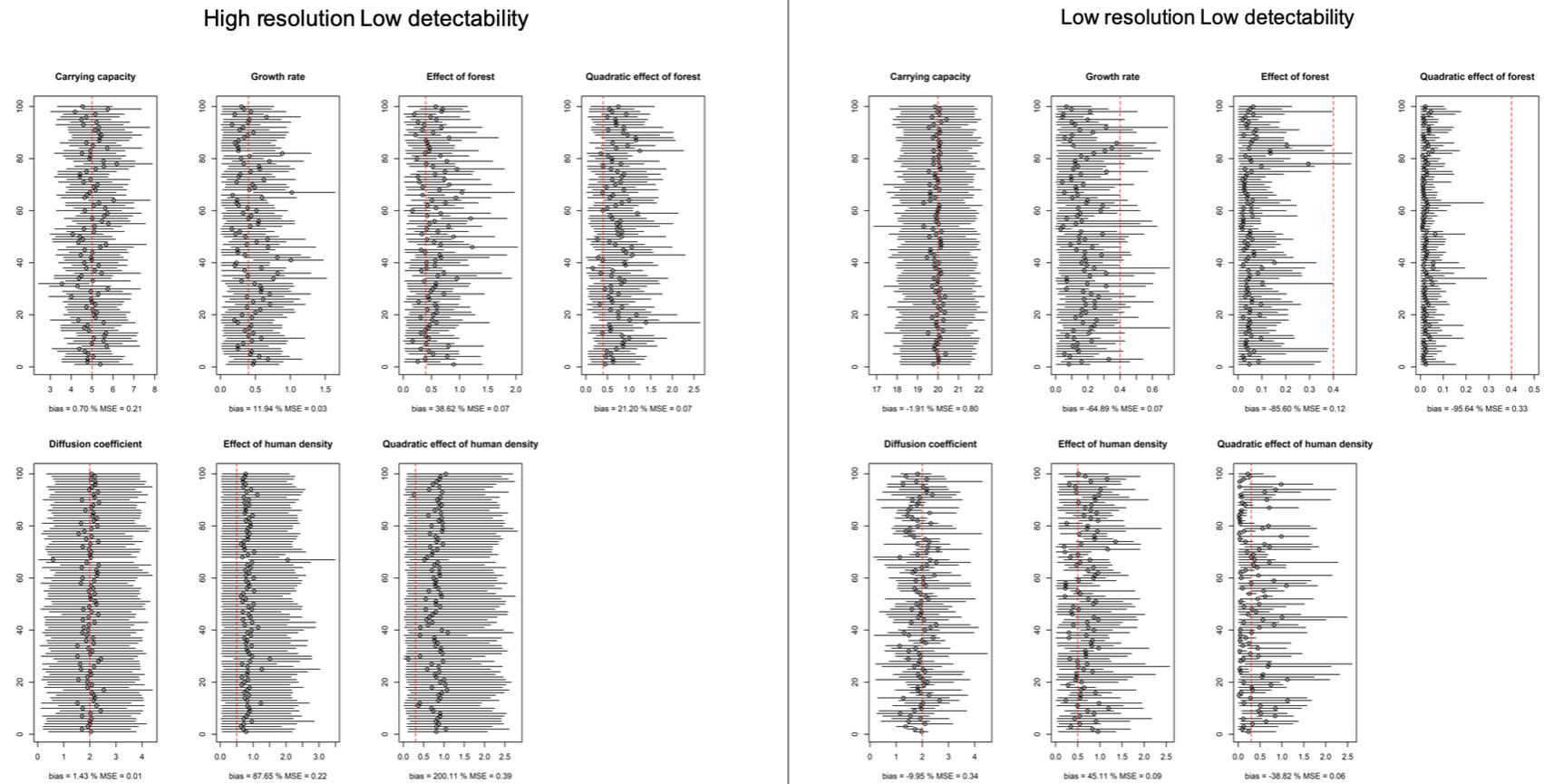

Appendix 2. B. Performance of the model without covariates in the high resolution / high detectability scenario (left panels) and in the low resolution / high detectability scenario (right panels). For each of the 100 simulated datasets (on the Y-axis), we displayed the median (circle) and the 95% credible interval (horizontal solid line) of the parameter. The actual value of the parameter is given by the vertical dashed red line. The estimated bias and MSE are provided in the legend of the X-axis.

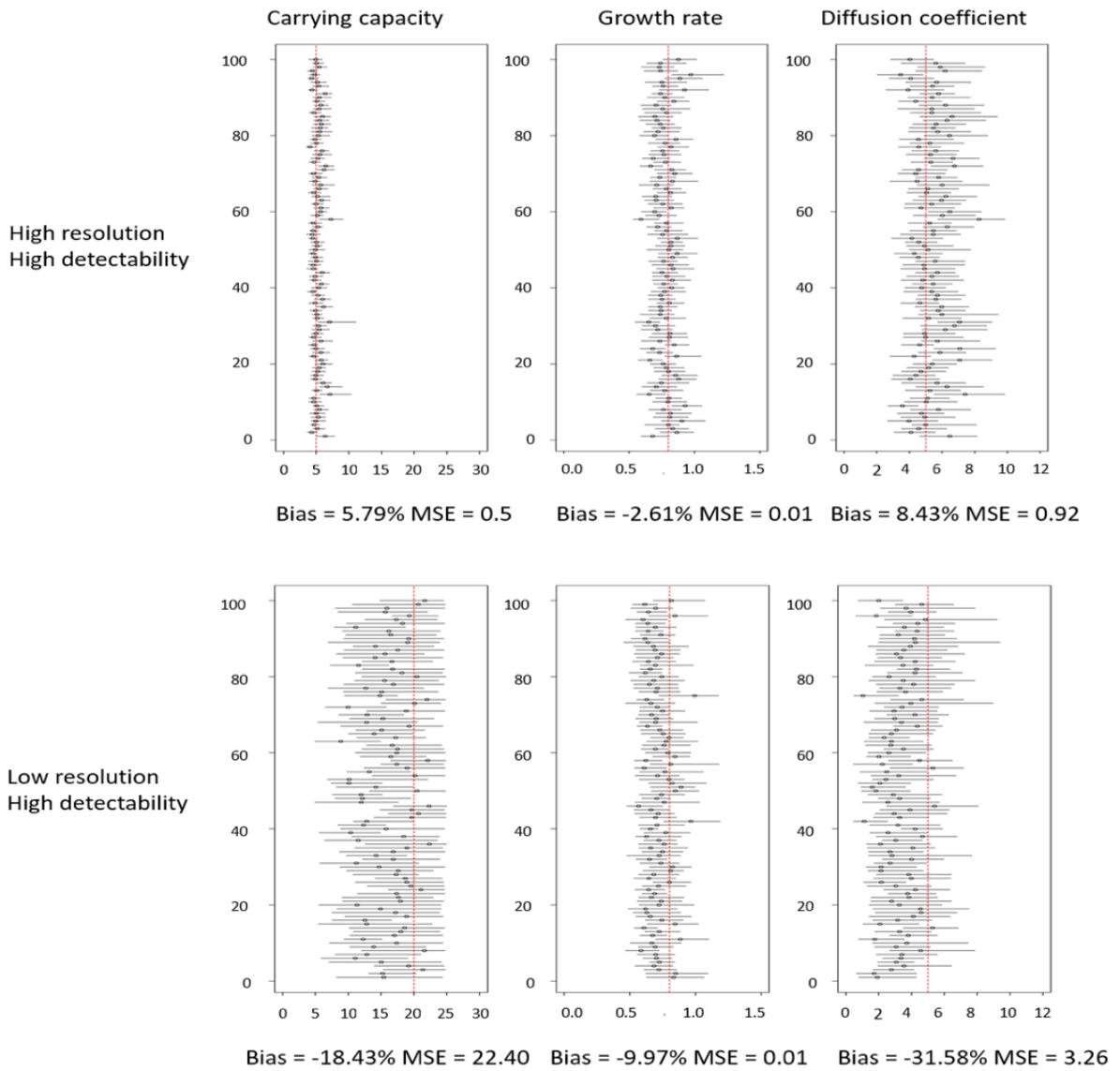

Appendix 2. C. Performance of the model without covariates in the high resolution / low detectability scenario (left panels) and in the low resolution / low detectability scenario (right panels). For each of the 100 simulated datasets (on the Y-axis), we displayed the median (circle) and the 95% credible interval (horizontal solid line) of the parameter. The actual value of the parameter is given by the vertical dashed red line. The estimated bias and MSE are provided in the legend of the X-axis.

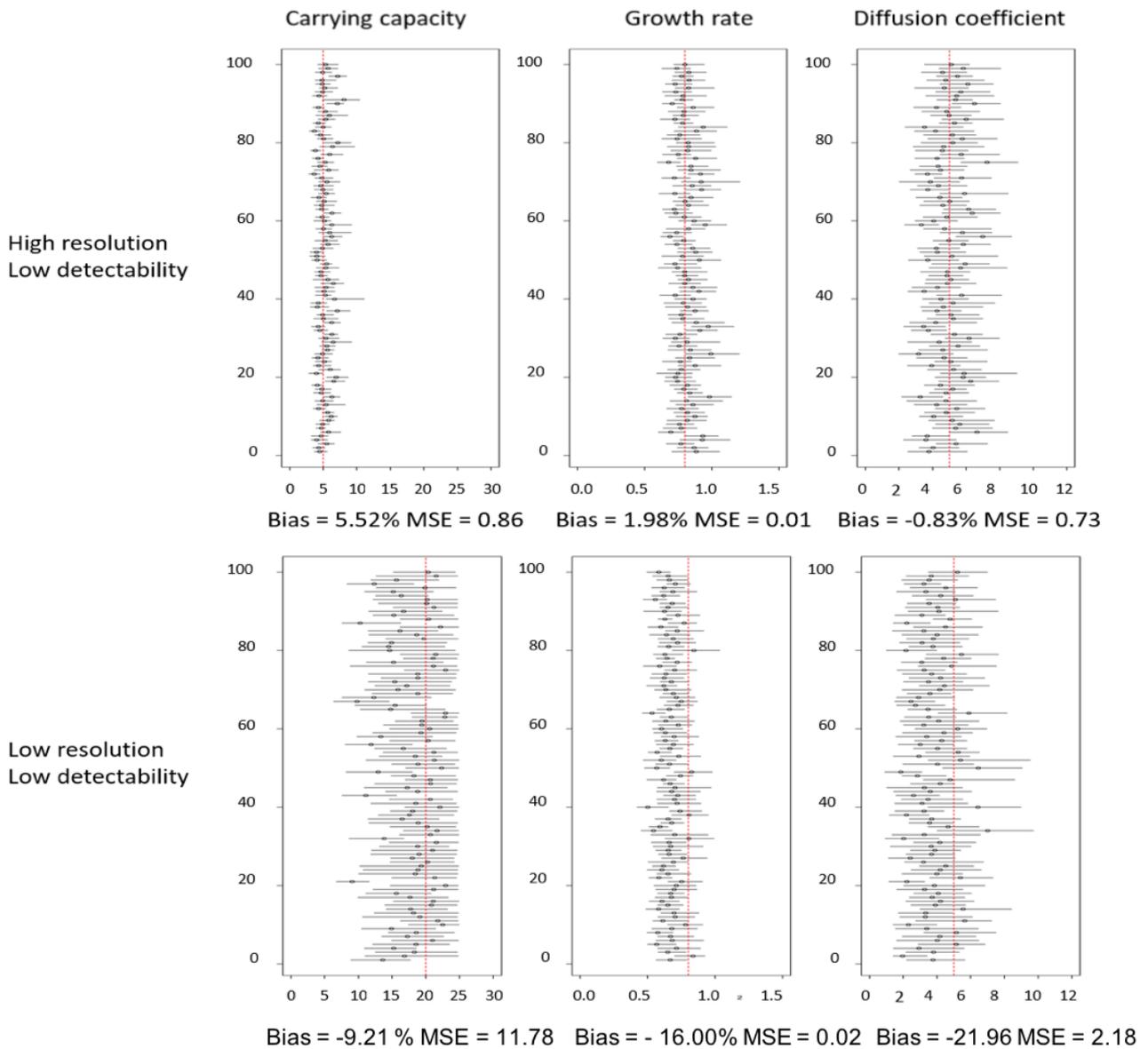

Appendix 3: Estimated abundance evolution for 10 years from the posterior median (red solid line) and the 95 % credible intervals (grey dashed line) in comparison with the true abundance (blue dashed line) for the first 25 sites in the two "high resolution" scenarios and the 25 sites in the two "low resolution" scenarios.

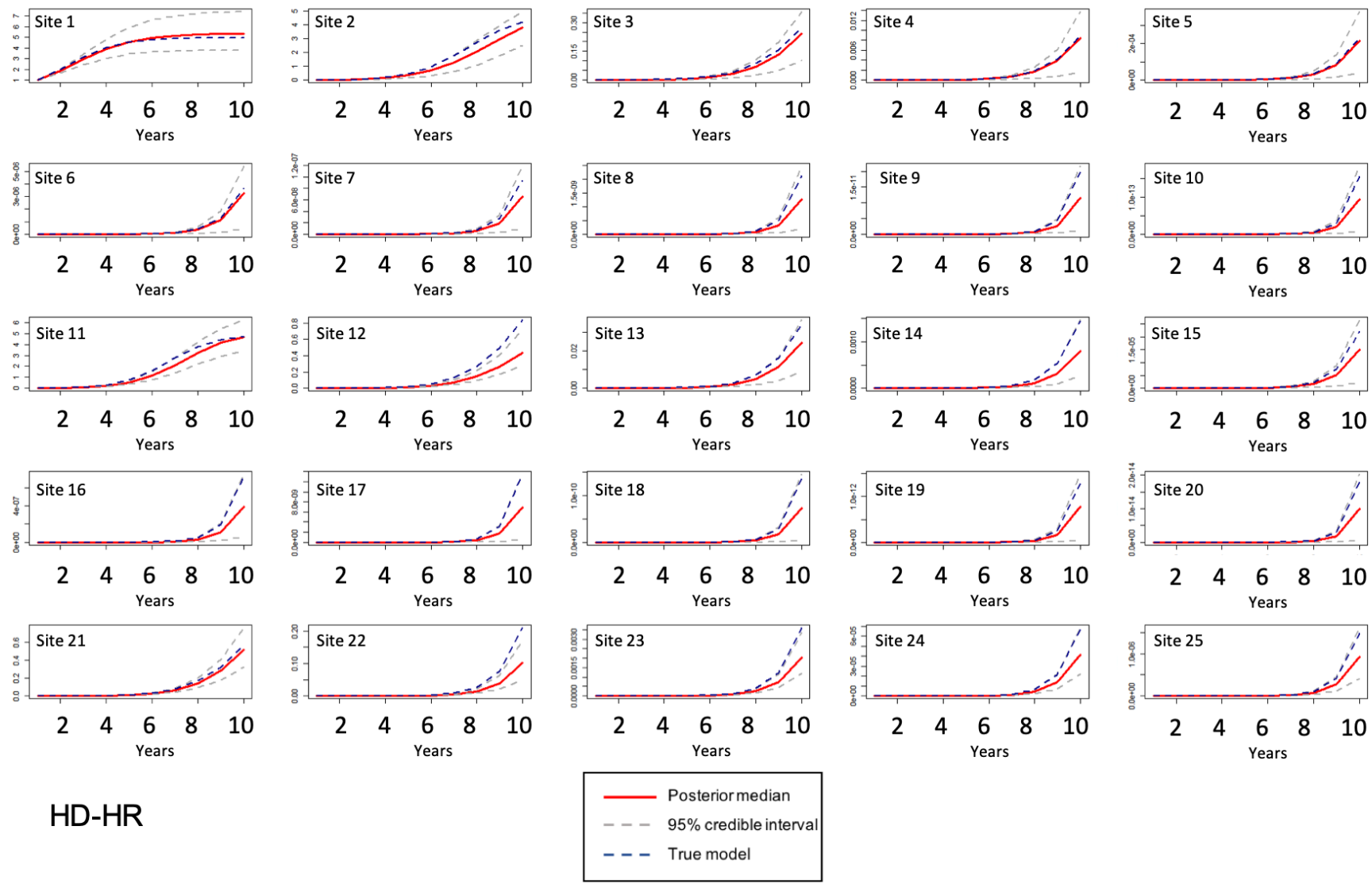

HD-HR

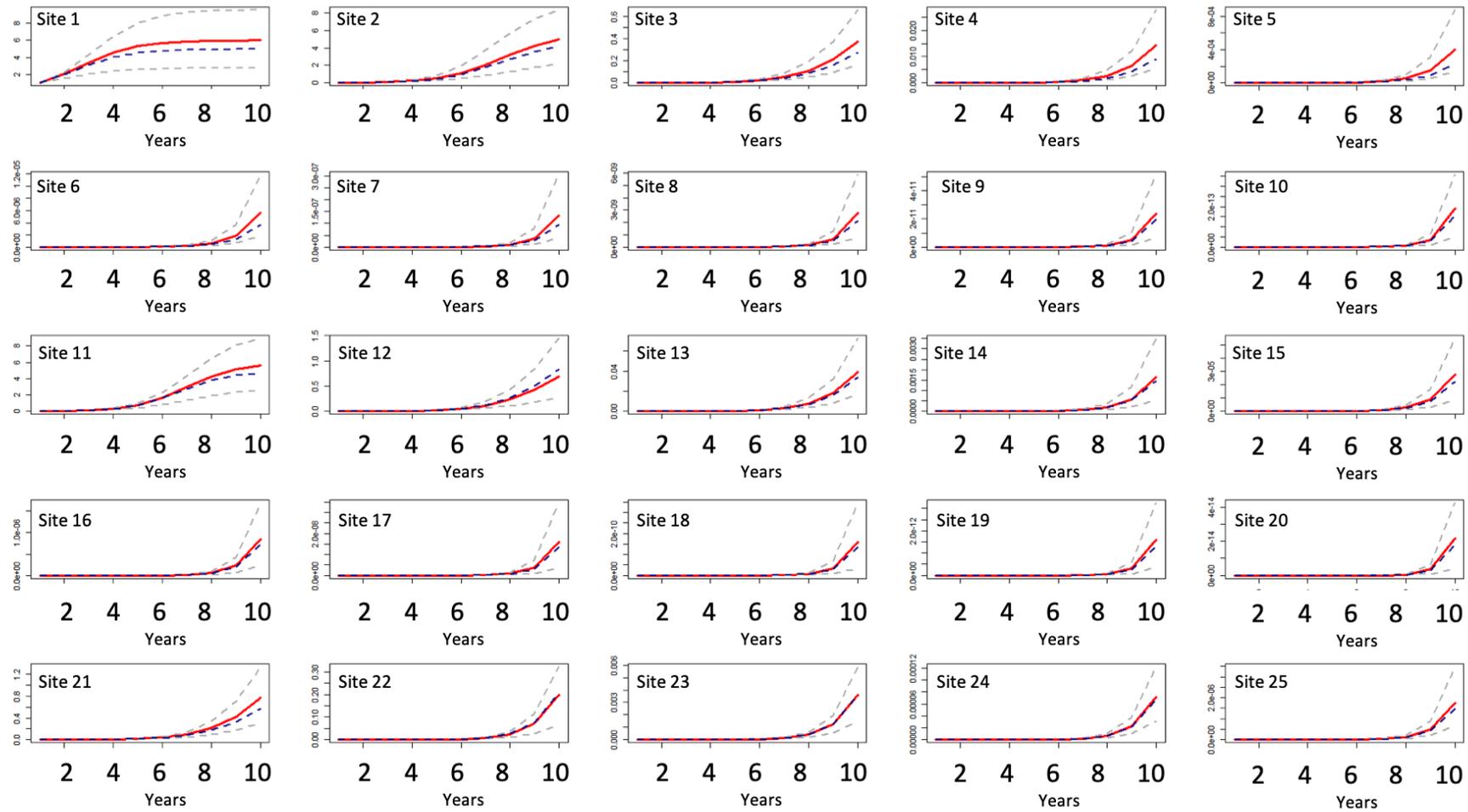

LD-HR

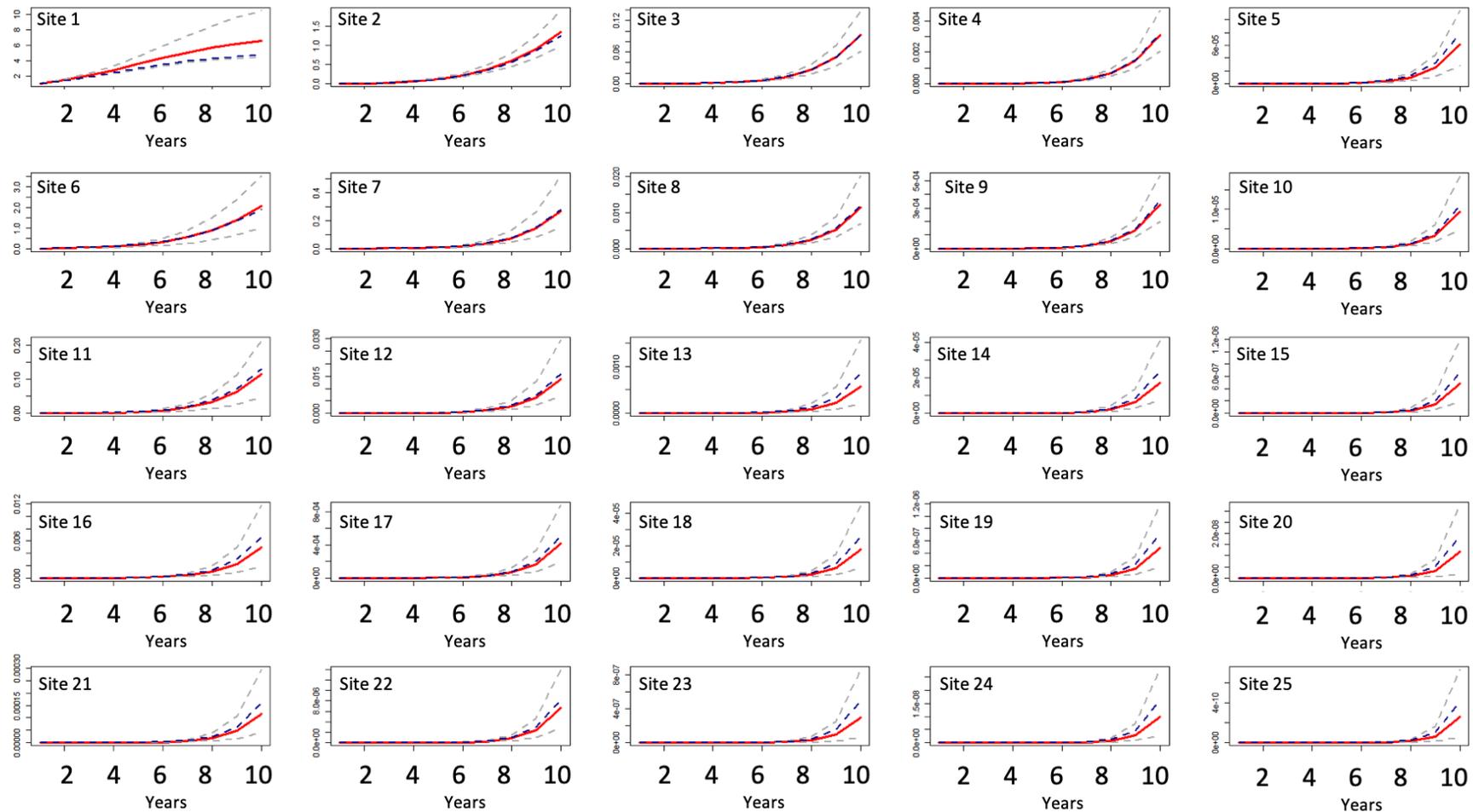

HD-LR

33
34

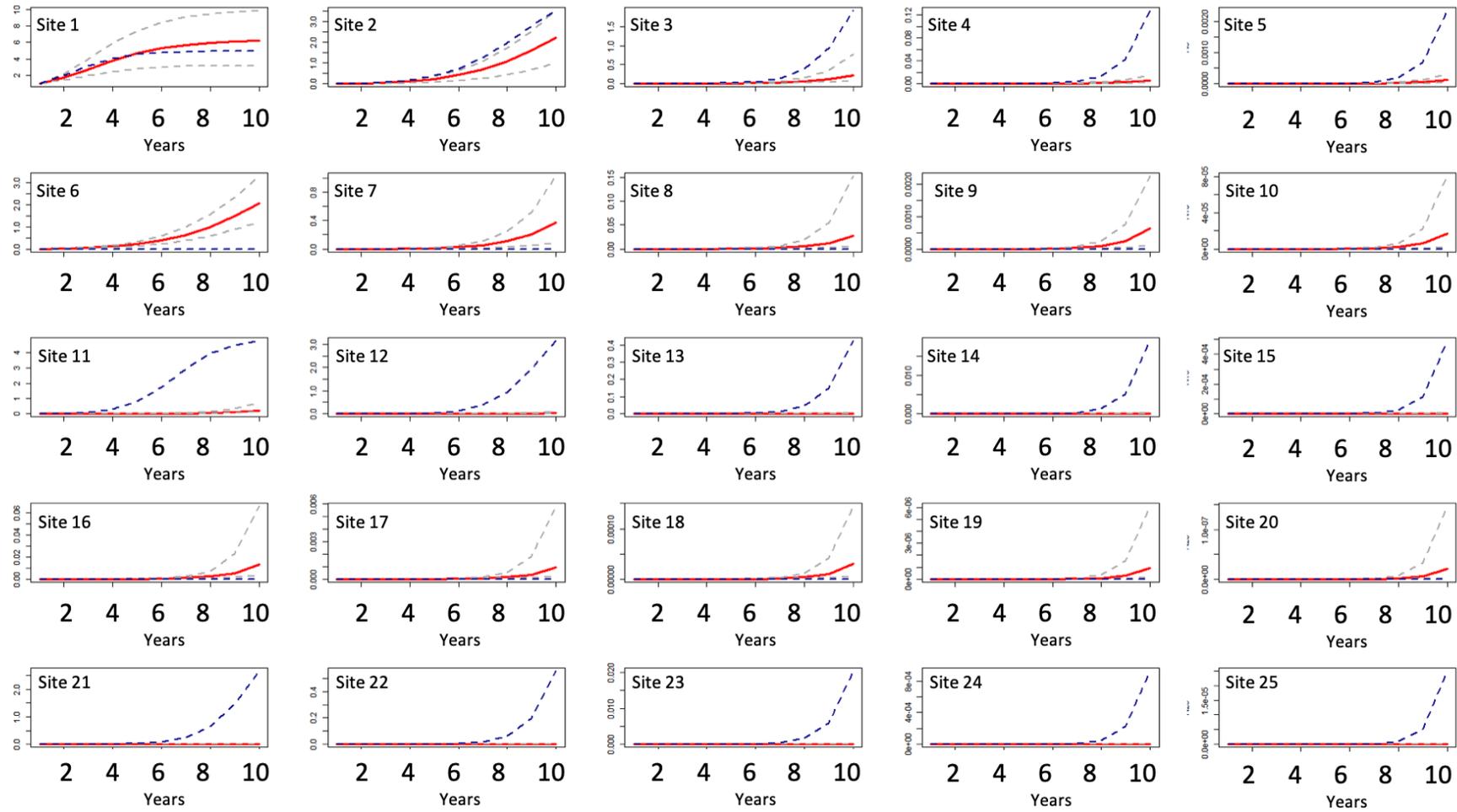

LD-LR

Appendix 4: Maps of the quantiles of the estimated abundance of wolves per site in South-East France between years 2007 and 2015. Black dots represent detections during a year.

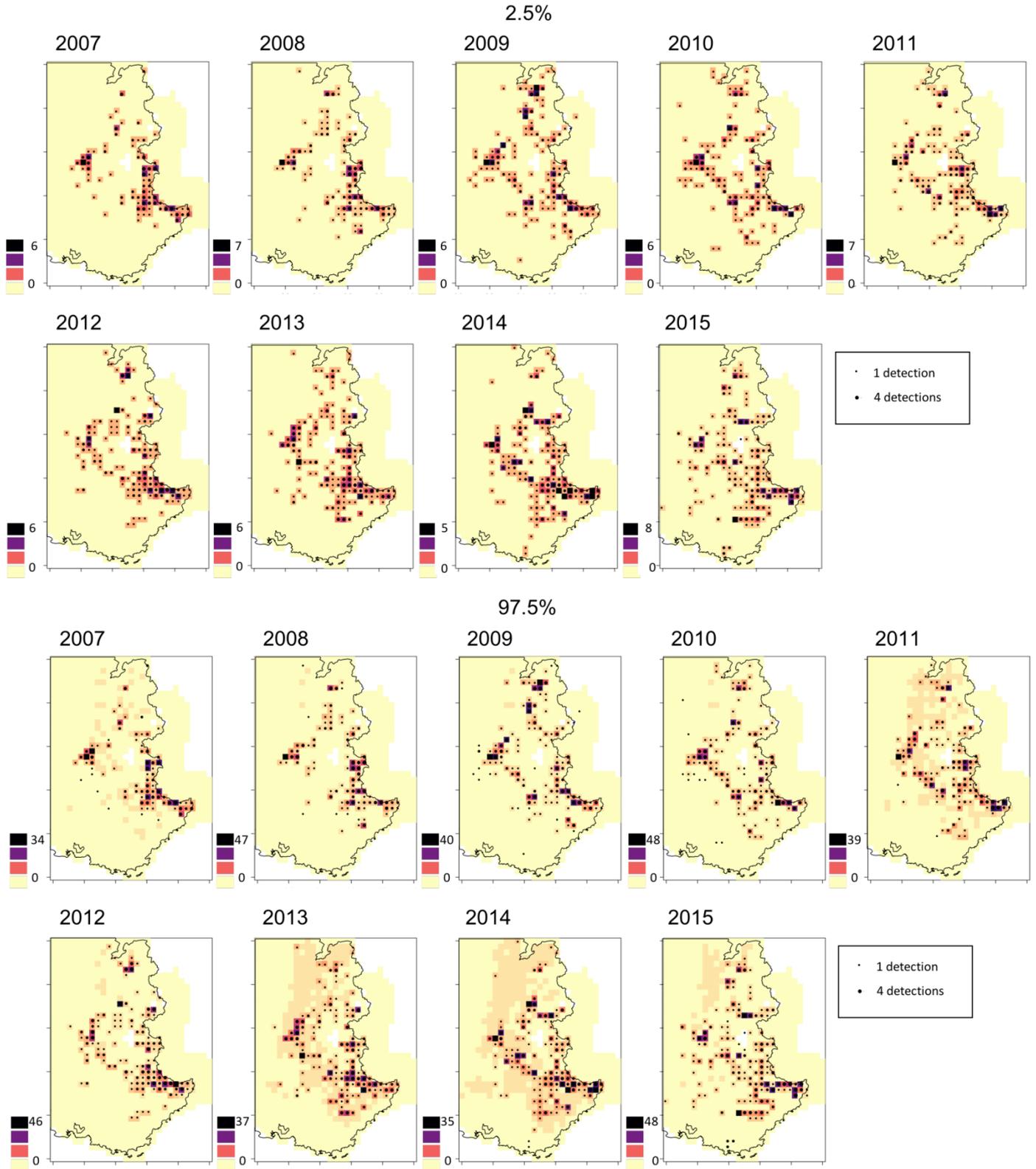

Appendix 5: Median and 95% credibility intervals for the parameters and the effects of ecological variables on wolf distribution dynamics between years 2007 and 2015 in South-Eastern France.

|  | 2.50% | 50% | 97.50% |
|---|---|---|---|
| **Species-level detectability $q$** | | | |
| Intercept | -2.83 | -2.59 | -2.30 |
| Linear effect of sampling effort | 0.21 | 0.34 | 0.45 |
| Quadratic effect of sampling effort | -0.85 | -0.71 | -0.59 |
| **Logistic growth rate $R$** | | | |
| Intercept | -0.47 | -0.44 | -0.41 |
| Linear effect of forest cover | 0.35 | 0.43 | 0.46 |
| Quadratic effect of forest cover | -0.47 | -0.44 | -0.32 |
| **Carrying capacity $K$** | | | |
| Intercept | $7.97 \times 10^{-3}$ | $9.41 \times 10^{-3}$ | $1.11 \times 10^{-2}$ |
| **Diffusion parameter $D$** | | | |
| Intercept | 0.92 | 1.25 | 1.55 |
| Linear effect of human density | 1.89 | 2.61 | 2.77 |
| Quadratic effect of human density | 0.11 | 1.26 | 2.11 |

Appendix 6: Maps of the quantiles, median and mean of the forecasted abundance of wolves per site in South-East France for 2016. Blue squares represent detections in year 2016.

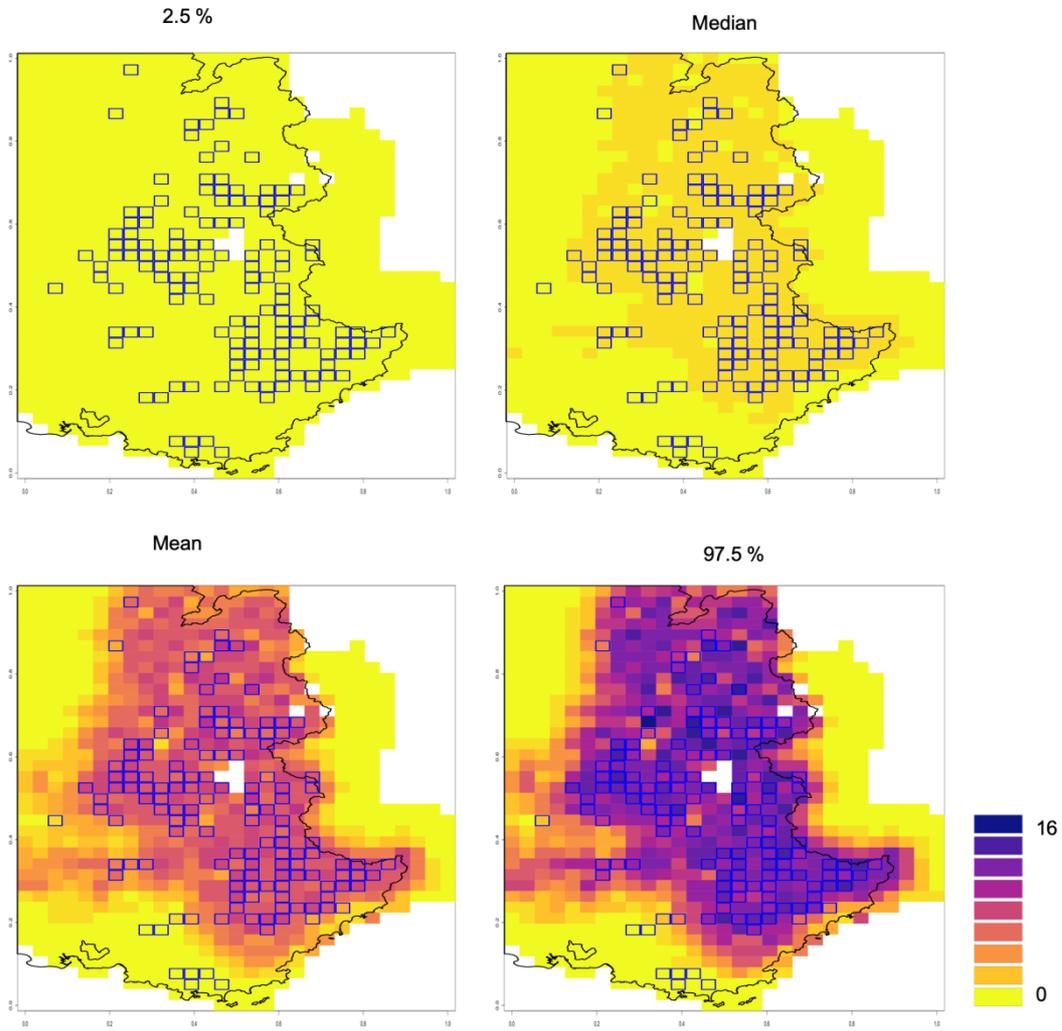